\documentclass[%
 reprint,
 amsmath,amssymb,
 aps,
 pra,
]{revtex4-2}

\usepackage{graphicx}
\usepackage{dcolumn}
\usepackage{bm}
\usepackage{color,soul}

\begin{document}

\title{4-field symmetry breakings in twin-resonator photonic isomers}
\author{Alekhya \surname{Ghosh$^{\;1,2}$}}
\author{Lewis \surname{Hill$^{\;1,3}$}}
\author{Gian-Luca \surname{Oppo$^{\;3}$}}
\author{Pascal \surname{Del'Haye$^{\;1,2,}$}}
\email{pascal.delhaye$@$mpl.mpg.de}
\affiliation{$^1$Max Planck Institute for the Science of Light, Staudtstra{\ss}e 2,
D-91058 Erlangen, Germany\\$^2$Department of Physics, Friedrich Alexander University Erlangen-Nurember, Staudtstra{\ss}e 2, D-91058 Erlangen, Germany\\$^3$Department of Physics, University of Strathclyde, 107 Rottenrow, Glasgow, G4 0NG, UK}

\begin{abstract}
Symmetry and symmetry breaking of light states play an important role in photonic integrated circuits and have recently attracted lots of research interest that is relevant to the manipulation of light polarisation, telecommunications, all optical computing, and more. We consider four-field symmetry breaking within two different configurations of photonic dimer systems, both comprised of two identical Kerr ring resonators. In each configuration we observe multiple degrees and levels of spontaneous symmetry breaking between circulating photon numbers and further, a wide range of oscillatory dynamics, such as chaos and multiple variations of periodic switching. These dynamics are of interest for optical data processing, optical memories, telecommunication systems and integrated photonic sensors.
\end{abstract}

\maketitle
\section{Introduction}
Spontaneous symmetry breaking (SSB) occurs when two or more properties of a system suddenly change from being equal (symmetric) to being unequal (asymmetric) following an infinitely small change to some system parameter. SSB phenomena have been found at the center of some of the most intriguing behaviors of physics~\cite{arodz2011patterns}, such as spontaneous breaking of gauge symmetry describing the Higgs mechanism~\cite{RevModPhys.46.7} and Einstein-Hilbert gravity in quantum field theory~\cite{RevModPhys.54.729}. Symmetry breaking has also been observed in two dimensional (2D) materials above Curie temperature~\cite{Liu2020} and leads to a large number of interesting applications in plasmonics~\cite{sym12060896}.

Over the last few decades, there have been many works looking to understand the behavior of high intensity light circulating in ring resonators made of nonlinear optical materials. This interest is based on their potential applications in telecommunications~\cite{kemal2020chip}, optical computing~\cite{Moroney:20}, metrology~\cite{pasquazi2018micro} and wider, and their ease of use for studying fundamental physical concepts, with the SSB of light being one of the most fruitful examples.

In particular the SSB of counter-propagating fields~\cite{KAPLAN1982229, PhysRevA.32.2857,PhysRevA.98.053863,PhysRevA.101.013823,DelBino2017,PhysRevLett.118.033901,PhysRevLett.126.043901, campbell2022counterpropagating} and the SSB of co-propagating orthogonally polarized fields~\cite{PhysRevLett.122.013905, geddes1994polarisation, garbin2020asymmetric} in Kerr ring resonators have led to many new applications. On the one hand, systems with counterpropagating light, initially proposed for enhancing the Sagnac effect~\cite{Kaplan:81, PhysRevA.32.2857, PhysRevA.98.053863, PhysRevA.101.013823, PhysRevLett.118.033901, campbell2022counterpropagating, PhysRevLett.126.043901}, can be used for isolators and circulators~\cite{DelBino:18}, logic gates~\cite{Moroney:20}, gyroscopes with enhanced sensitivity~\cite{Silver:21} and near field sensors, while on the other hand, the symmetry breaking between different polarizations has seen application in the production of vector solitons and breathers~\cite{Xu:22}, polarization controllers~\cite{Moroney2022} and even random number generators~\cite{Quinn:23}. SSB of solitons in Fabry Perot resonators has been recently reported~\cite{hill2023symmetry}.

A comparatively novel method of achieving SSB in Kerr ring resonators, which also serves as inspiration for this current work, is through the exploitation of identical, or ``twin" ring resonators~\cite{PhysRevA.95.053822}. SSB was recently observed in an evanescently-coupled Bose-Hubbard dimer where the intracavity photons experience a Kerr-like optical nonlinearity~\cite{PhysRevLett.128.053901}. 
 By observing not one, but two, twin resonator systems, and considering polarization effects, we describe methods of achieving highly controllable multi-staged SSB with a wide range of different field dynamics, such as, oscillatory, chaotic and self-switching.

An enormous benefit of twin-resonator systems studied here over a recently reported alternative multi-staged SSB system~\cite{hill2023multi} lies in its degree of controllability, thus giving increased freedom and flexibility for fundamental science experiments and applications.

\begin{figure}[t]
\includegraphics[width=1\columnwidth]{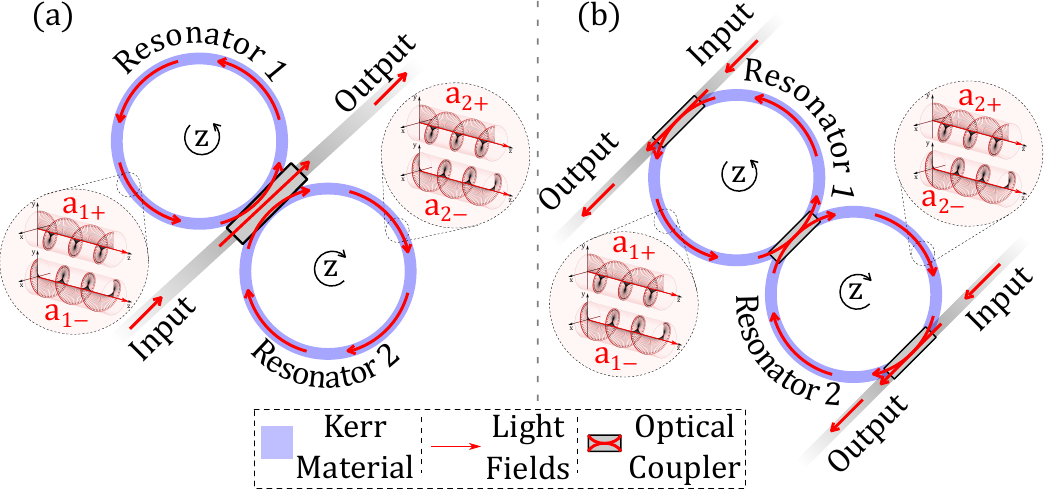}
\caption {\textit{Photonic dimer configurations.} Two identical Kerr ring resonators receive identical, linearly polarized, input beams inducing circulating fields within the two resonators. By considering field polarization we model a total of four circulating fields, represented by modal amplitudes of the field components $a_{1\pm,2\pm}$. The circulating fields of the two twin resonators can exchange power through appropriate means, such as a fiber coupler or evanescent field coupling. In system ``o$\arrowvert$o", (a), the fields within the two resonators are connected through the waveguide between said resonators and in system ``$\arrowvert$oo$\arrowvert$", (b), the fields within the two resonators are geometrically overlapping.}
\label{fig1}
\end{figure}

We present in Fig. \ref{fig1} the schematics of our two systems of study. We shall refer to these configurations by the names ``o$\arrowvert$o" (pronounced ``olo") and ``$\arrowvert$oo$\arrowvert$" (pronounced ``lool") respectively. Even visually, one can see that while there are many similarities between the two systems, there are also some key differences between them. In both systems, the Kerr ring resonators are modeled as perfect copies of each other, or ``twins", where linearly polarized light is coupled into the resonators by inputs, and where both fields within the resonators are projected onto left- and right- circular polarisation components. The mechanism of field cross-talk between the resonators in the two systems differ. In system ``o$\arrowvert$o", Fig. \ref{fig1}a, the two resonators are not geometrically coupled to each other, but are instead symmetrically coupled to and by a single common input channel positioned between them, which further provides linearly polarised light to both resonators symmetrically. In system ``$\arrowvert$oo$\arrowvert$", Fig. \ref{fig1}b, the resonators are instead directly coupled to each other forming a photonic dimer, and are further uniformly coupled to two separate input channels, each providing linearly polarised light of matching intensity, frequency and polarization direction to the resonators. For understanding the implications of these differences between the systems, it is important to note that in system ``$\arrowvert$oo$\arrowvert$" there is direct geometrical overlap between the fields circulating the two resonators, whereas in system ``o$\arrowvert$o" the distance between the resonators is such that this overlap does not exist. In system ``o$\arrowvert$o" however the fields that comes out of one resonator can still enter the other, only this time via the intermediary channel.

\section{Model}
For modeling the resonator systems we start with base equations from Ref.~\cite{Tikan2021, PhysRevA.95.053822} and add additional terms that describe the Kerr nonlinearity. A detailed derivation can be found in Appendix A. We consider

\begin{equation}
    \begin{split}
        \dot{a}_{1\pm,2\pm} = \left(i\Delta -\frac{\kappa}{2}\right) &a_{1\pm,2\pm} +\zeta a_{2\pm,1\pm} + i U |a_{1\pm,2\pm}|^2 a_{1\pm,2\pm}\\
        &+ i 2U |a_{1\mp,2\mp}|^2 a_{1\pm,2\pm} + \sqrt{\kappa_e}S_{in},
    \end{split}
    \label{LangevinEquations}
\end{equation}

\noindent where $\Delta = \omega_0 - \omega_l$ is the cavity detuning (the difference between the input frequency and the closest cavity resonance frequency), $\kappa = \kappa_l + \kappa_e$ is the total loss, with internal losses $\kappa_l$ and external losses $\kappa_e$. The term $\zeta$ describes the coupling mechanism between the two resonators and is given by
\begin{subequations}
\begin{eqnarray}
\zeta & = & +iJ, \text{  for system ``$\arrowvert$oo$\arrowvert$"},\\
&=& -\frac{\kappa_e}{2}, \text{  for system ``o$\arrowvert$o"},
\label{zetaExp}
\end{eqnarray}
\end{subequations}
where $J$ is the coupling rate between the two resonators in system ``o$\arrowvert$o"~\cite{Tikan2021}. The fourth and fifth terms of Eq.~\eqref{LangevinEquations} are self- and cross phase modulation terms, respectively, which account for the nonlinear effects of a field on itself and of other fields on the equations primary field, respectively, with $U = \frac{\hbar \omega_0^2 c n_2}{n_0^2 V_{eff}}$ being the Kerr coefficient, where $c$ is the speed of light, and $n_2$ and $n_0$ are respectively the nonlinear and linear refractive indices of the medium. The final term of Eq.~\eqref{LangevinEquations} represents input from outside the system, where $|S_{in}|^2$ is the input photon flux. Since the two ring resonators in both cases are identical, parameters such as the cavity detuning and the Kerr-nonlinear coefficients, $U$, are the same for both resonators. We consider group-velocity dispersion to be negligible in this work.

\begin{figure}[b!]
\includegraphics[width=1\columnwidth]{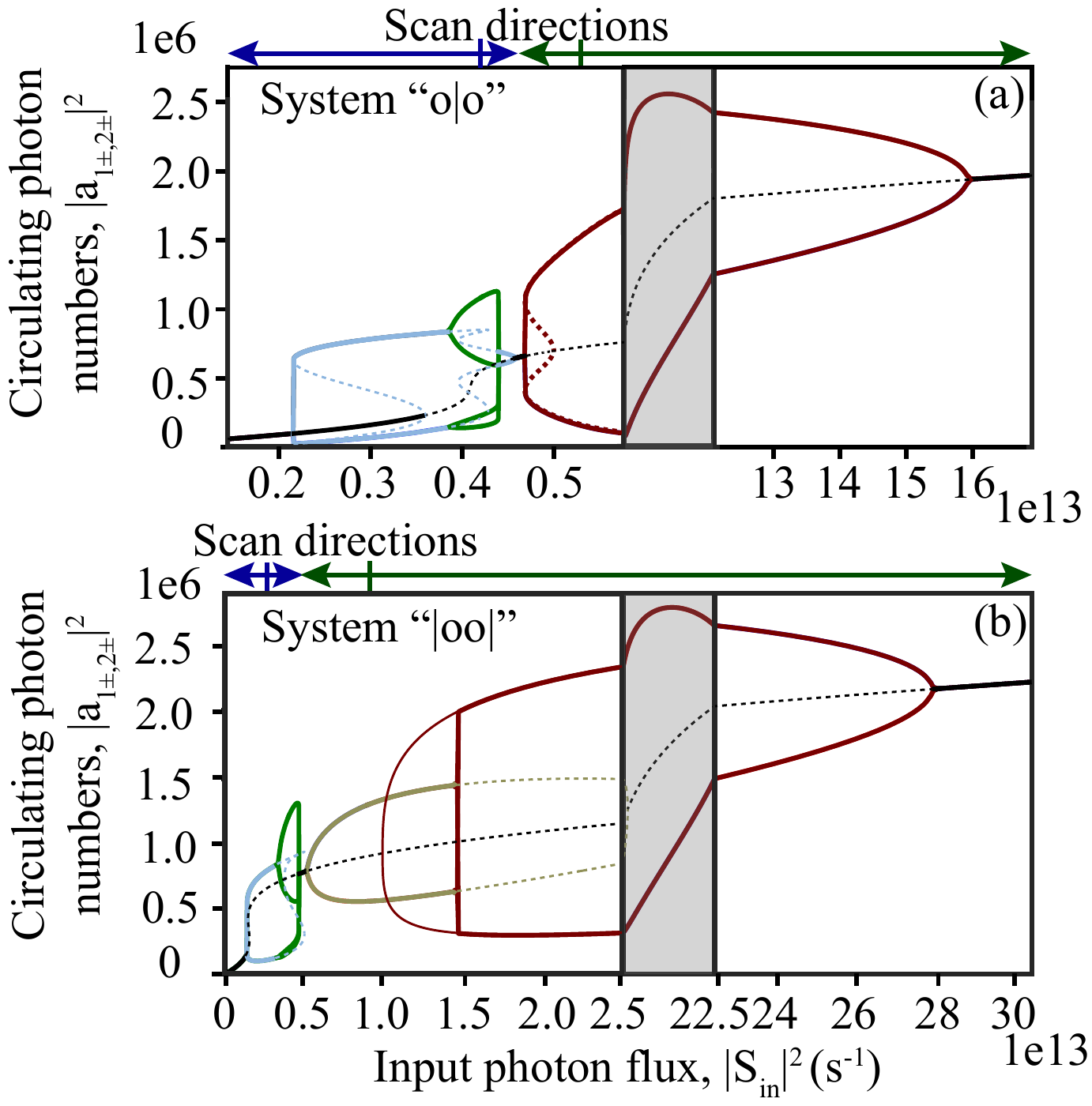}
\caption {Input intensity scans. (a) and (b) show the variations of the circulating intracavity photon numbers $|a_{1\pm,2\pm}|^2$ for the ``o$\arrowvert$o" and ``$\arrowvert$oo$\arrowvert$" systems respectively; obtained using Eq.~\eqref{LangevinEquations} for a cavity detuning of $\Delta = -2.52 \kappa_l$ for (a) and $\Delta = -2.45 \kappa_l$ for (b). The bold lines show the results of simulation. The pale solid and pale dashed lines show the stable and unstable regions of the analytical solution, respectively. The black lines represent the fully symmetric solution, the sky-blue lines represent the Polarization Symmetry bubble, the green lines show the fully asymmetric bubble, the brown lines depict the Cross Symmetry bubble and the yellow lines show the Resonator Symmetry bubble, with these symmetries defined in Table~\ref{tab:my_label}. The scan directions for both cases are shown above the plots. The long monotonous region in the Cross Symmetry bubble has been squeezed in the gray region.
 Used parameters: $\kappa_e = \kappa_l = \pi$ MHz, $U = 4$. For all the simulations in this work we have considered $J = \kappa_e/2$.}
\label{fig2}
\end{figure}

\begin{figure*}
\includegraphics[width=0.75\textwidth]{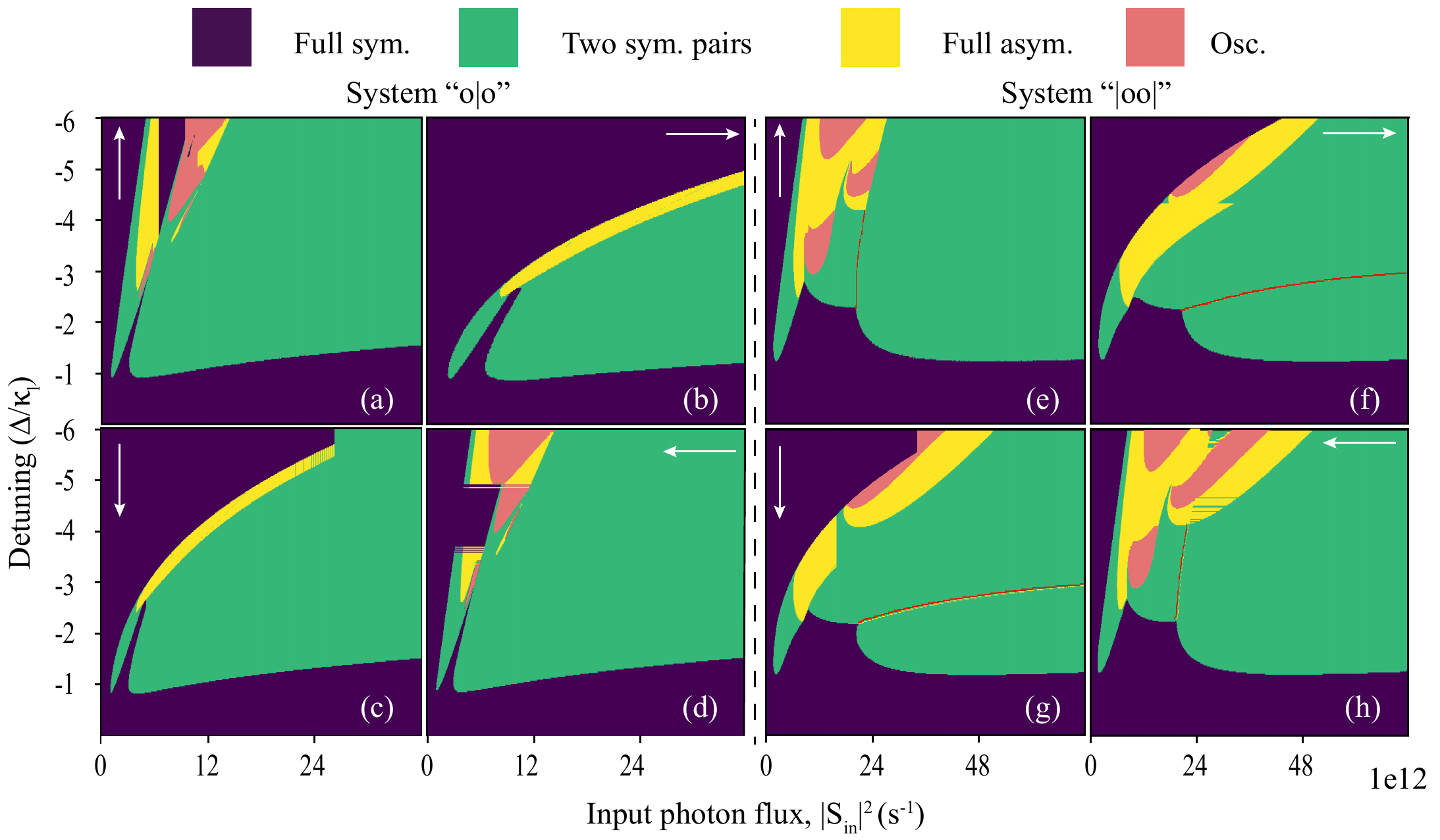}
\caption {\textit{Input power - detuning parameter scan}. (a,b,c,d) scans for system ``o$\arrowvert$o"; (e,f,g,h) scans for system ``$\arrowvert$oo$\arrowvert$". Purple corresponds to regions with symmetric field intensities. Green corresponds to regions with a single symmetry breaking bubble (RSB or PSB or XSB), i.e., two pairs of symmetric fields are different from each other in this area. Yellow shows where all the four fields are different (fully asymmetric). Oscillations in the field intensities can be observed in the pale red zones. The dark red lines in (e,f,g,h) denote the small 4D-oscillatory segments during transitions from PSB to SSB regions. The white arrows show the directions of the scans (e.g. arrow up = increasing detuning; arrow left = decreasing input power). For these simulations we use $U = 4$.}
\label{fig4}
\end{figure*}
\section{Sequential and nested SSB}
We begin by seeking the set of stationary states of Eq.~\eqref{LangevinEquations}, where the fields $a_{1\pm,2\pm}$ do not change over time, i.e. $\dot{a}_{1\pm,2\pm}=0$. We can find analytically a partial set of these stationary states by forcing degeneracies, or symmetries, on the system (such as forcing $a_{1+,2+}=a_{1-,2-}$, detailed calculations are provided
in the Appendix). For the system with no such forced symmetries however, we numerically evaluate Eq.~\eqref{LangevinEquations} for a variety of system parameters, and over sufficient evolution times to find additional stationary states. The initial condition for the zero-input power is defined as all the four field components have zero amplitudes and zero phases. Thereafter, to replicate the experimental conditions, where the input power is increased continuously at a rate much slower than the cavity build-up time, the system is allowed to evolve for a time much longer than the cavity build-up time, and after it reaches steady state, that steady state values of the field components are used as the initial condition for the evolution of the system with the next input power. The time step for integration is considered to be 5 ns, the total integration time for achieving steady state is considered to be greater than 60 $\mu$s, the step-size of increment of $S_{in}$ is from $(1.2 \sim 2.68)\times10^4$. The step-size is chosen to be big (small) in regions where the changes in the steady-state amplitudes of the circulating fields are small (big). Figure \ref{fig2} shows the results of this analysis in the form of input intensity scans.

\begin{table}[]
    \centering
    \begin{tabular}{c|c}
        \textbf{Degree of Symmetry}\; & \;\textbf{Fields' Intensity Relation} \\[1ex] 
        \hline\\[-1ex] 
        Full Symmetry\; & \; $|{a}_{1+}|^2 = |{a}_{1-}|^2 = |{a}_{2+}|^2 = |{a}_{2-}|^2$ \\
        Polarization Symmetry\; & \; $|{a}_{1+}|^2 = |{a}_{1-}|^2 \neq |{a}_{2+}|^2 = |{a}_{2-}|^2$\\
        Resonator Symmetry\; & \; $|{a}_{1+}|^2 = |{a}_{2+}|^2 \neq |{a}_{1-}|^2 = |{a}_{2-}|^2$ \\
        Cross Symmetry\; & \; $|{a}_{1+}|^2 = |{a}_{2-}|^2 \neq |{a}_{1-}|^2 = |{a}_{2+}|^2$ \\
        Full Asymmetry\; & \; $|{a}_{1+}|^2 \neq |{a}_{1-}|^2 \neq |{a}_{2+}|^2 \neq |{a}_{2-}|^2$\\[-1ex] \;
    \end{tabular}
    \caption{The circulating photon number relations that correspond to various stages and types of SB in our systems.}
    \label{tab:my_label}
\end{table}

From Fig.~\ref{fig2}, it can be seen that for small input powers all the four fields are symmetric in their intensities, defined in the first line of Table~\ref{tab:my_label}. When we define the system as holding full symmetry between the circulating photon numbers, the system holds all the following symmetries and corresponding invariances Polarization Symmetry (PS): $|a_{1+}|^2\leftrightarrow|a_{1-}|^2 \;\&\; |a_{2+}|^2\leftrightarrow|a_{2-}|^2$, Resonator Symmetry (RS): $|a_{1+}|^2\leftrightarrow|a_{2+}|^2 \;\&\; |a_{1-}|^2\leftrightarrow|a_{2-}|^2$, and Cross Symmetry (CS): $|a_{1+}|^2\leftrightarrow|a_{2-}|^2 \;\&\; |a_{1-}|^2\leftrightarrow|a_{2+}|^2$.

After a certain threshold, which is highly dependent on system parameters, this full symmetry is partially lost, and the fields separate into two stable asymmetric pairs of symmetric fields (blue solid lines in Fig.~\ref{fig2}a $\&$ c). In keeping with convention, we refer to this point of partial symmetry loss as a SSB bifurcation. 
At this SSB bifurcation, the fields are forced 
to pair up with symmetric polarisation components within each resonator, Table~\ref{tab:my_label} row 2, which amounts to the effect of both the Resonator and Cross Symmetries Breaking (RSB \& CSB respectively). RSB refers to the situation when one resonator's total intensity is suppressed and the other's is enhanced, while CSB means that the symmetry which used to exist between the intensities of the right-circularly polarized component of one resonator and the left-circularly polarised component of the other has broken.

Above a second input power threshold, it can be seen that each of the two pairs of symmetric fields experience a second SSB bifurcation, where the final symmetry, the polarization symmetry, also breaks -- resulting in the system having full asymmetry between the circulating photon numbers, Table~\ref{tab:my_label} row 5.

The inverse bifurcations of the fully asymmetric regions can then be observed in both systems, where various symmetries are restored until again the four fields behave symmetrically for a small range of input intensities.

Continuing to observe Fig.~\ref{fig2} for even higher input powers, alternative SSB bifurcations occurs for both systems. The respective symmetries that break at each SSB bifurcation are different for the two systems. In system ``o$\arrowvert$o", the SSB leads to the field pairing with our previously defined Cross Symmetry, Table~\ref{tab:my_label} row 4. However, in system ``$\arrowvert$oo$\arrowvert$", two distinct SSB bubbles occur, each with their own unique SSB bifurcations. The first bifurcation breaks both PS and CS and leads to the field pairing with  RS alone as shown in the third row of Table~\ref{tab:my_label}. This RS pairing has not been observed in system ``o$\arrowvert$o".

The steady state solutions later jump to other steady state solutions mid-RS bubble. These solutions correspond to the field pairings with CS. This jump of the system state from one stable condition to another stable condition is a particularly interesting feature of the system ``$\arrowvert$oo$\arrowvert$". By close inspection of the crossing point, it has been observed that the PSB bubble breaks into a set of fully asymmetric solutions where the four fields start to oscillate. The oscillations trigger the system to change the state.If the two resonators in system ``$\arrowvert$oo$\arrowvert$" or ``o$\arrowvert$o" are assumed to be different, since the symmetry between the resonators is not present, one expects for example the pitchfork bifurcations of Fig.~\ref{fig2} to transform into saddle-node bifurcations that is typical of imperfect bifurcations of this kind~\cite{Golubitski79, Golubitsky85}.

\begin{figure}[b]
\includegraphics[width=1\columnwidth]{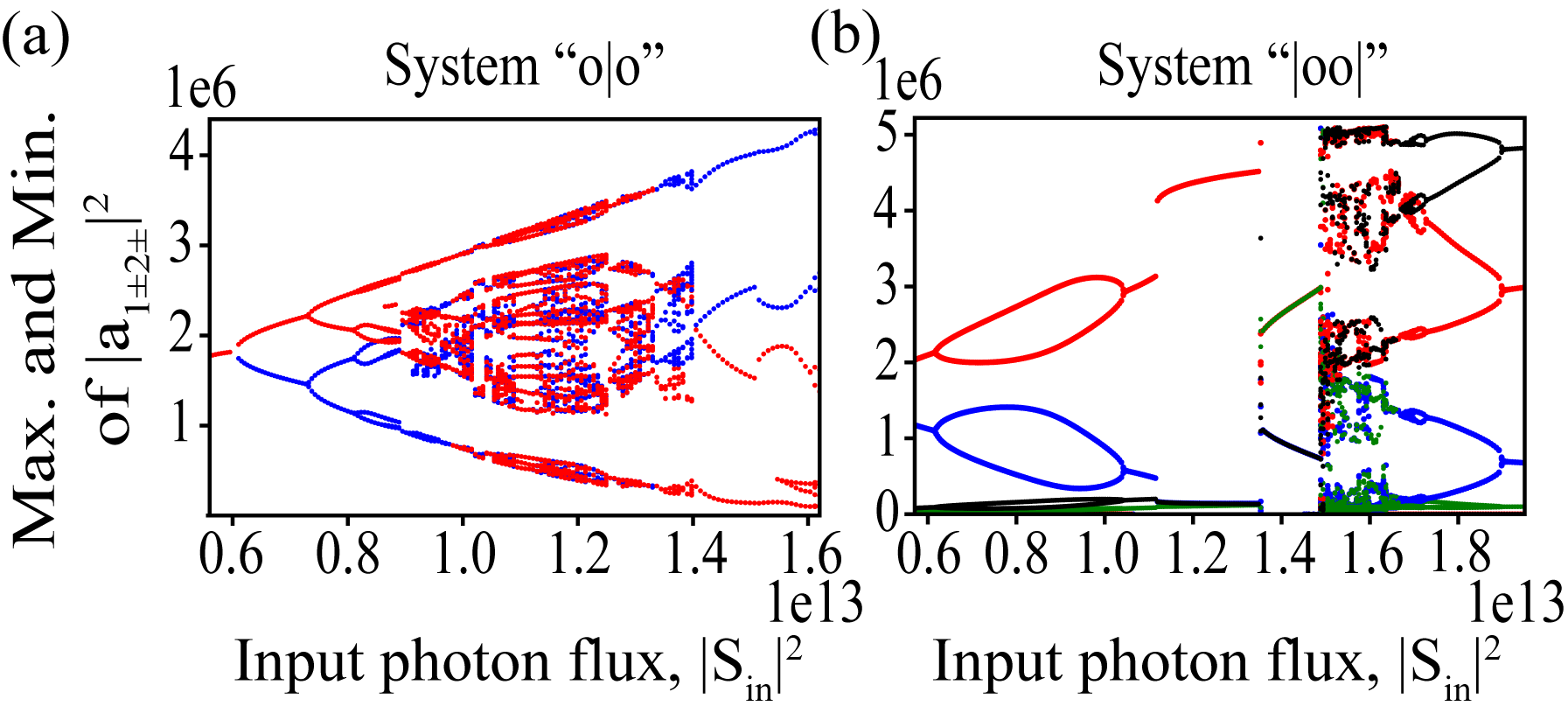}
\caption {\textit{Poincare sections of oscillations in the system}. Maxima and minima of the four field components for system ``o$\arrowvert$o" (a) and system ``$\arrowvert$oo$\arrowvert$"(b). The maxima and minima of $|a_{1+}|^2$ are shown by blue dots, of $|a_{1-}|^2$ by red dots, of $|a_{2+}|^2$  by black dots and of $|a_{2-}|^2$ by green dots. A single point of a particular color for a certain input power indicates the absence of oscillation for that field, whereas, two points at a given input power correspond to oscillations and a lot of such points refer to chaos. In (a), for lower input power, the system exhibits no oscillations and $|a_{1+}|^2 = |a_{1-}|^2$ . The first bifurcation of red and blue lines shows a SSB between the two fields, whereas, the bifurcation of the single red/blue line to two red/blue lines depict oscillations in the system, the amplitude of which is bounded by the two red/blue lines. The oscillations then overlap and lead to chaos. In (b) Uncoupled oscillations in all the fields appear for lower input power followed by regions of 4D and 2D SSB. Chaos in (b) is indicated by complete overlap of oscillations of the four fields. The chaos ends with uncoupled oscillations of the four fields towards higher input power, which further lead to a 4-D SSB region without any oscillations.}
\label{fig5}
\end{figure}

\section{Parameter scans}
To deepen our understanding of the SSB behaviours within the system described by Eq.~\eqref{LangevinEquations}, we show in Fig.~\ref{fig4} parameter space scans for the two systems over the input intensity and cavity detuning parameters, where we further scan from various directions to capture different possibilities of bistable system states. Within these scans, we not only show the varying degrees of symmetry between the circulating photon numbers, but also where the photon numbers show \textit{oscillatory behaviour}.

From Fig.~\ref{fig4}, it is evident that both systems can exhibit oscillatory behavior for certain ranges of values for input power and detuning. Different distinct regions in the parameter scan regions correlate to different types of oscillations, often with different pairings of the fields and their relative phases. One method to visualize the oscillations and the presence of chaos in a system is to generate Poincar\'e section plots. In Fig.~\ref{fig5}, Poincar\'e sections at the maxima and minima of the field intensities for the two systems are presented. For the first system (Fig~\ref{fig5}a), the maxima and minima of the two dominant fields have been plotted for a detuning $\Delta = -6.7 \kappa_l$. With increasing power, at first the symmetry between the two fields breaks and thereafter the fields start to oscillate. The maxima and minima of the fields diverge with increasing power, and after a small region of cascading period doubling bifurcations, the maxima of the lower field cross the minima of the upper field causing a region of overlap. This begins a region of chaotic oscillations. After the chaotic region, the oscillations of the two fields decouple and the system returns to a more regular form of oscillatory behaviour. In the Poincar\'e section of the second system (Fig~\ref{fig5}b) for a detuning $\Delta = -5.82 \kappa_l$, decoupled symmetry broken oscillations of the four fields emerge from the fully symmetry broken condition at the beginning of the plot. From an input flux of $|S_{in}|^2 = 1.5\times 10^{13}$ a short region of chaos is observed.

\begin{figure}[h]
\includegraphics[width=1\columnwidth]{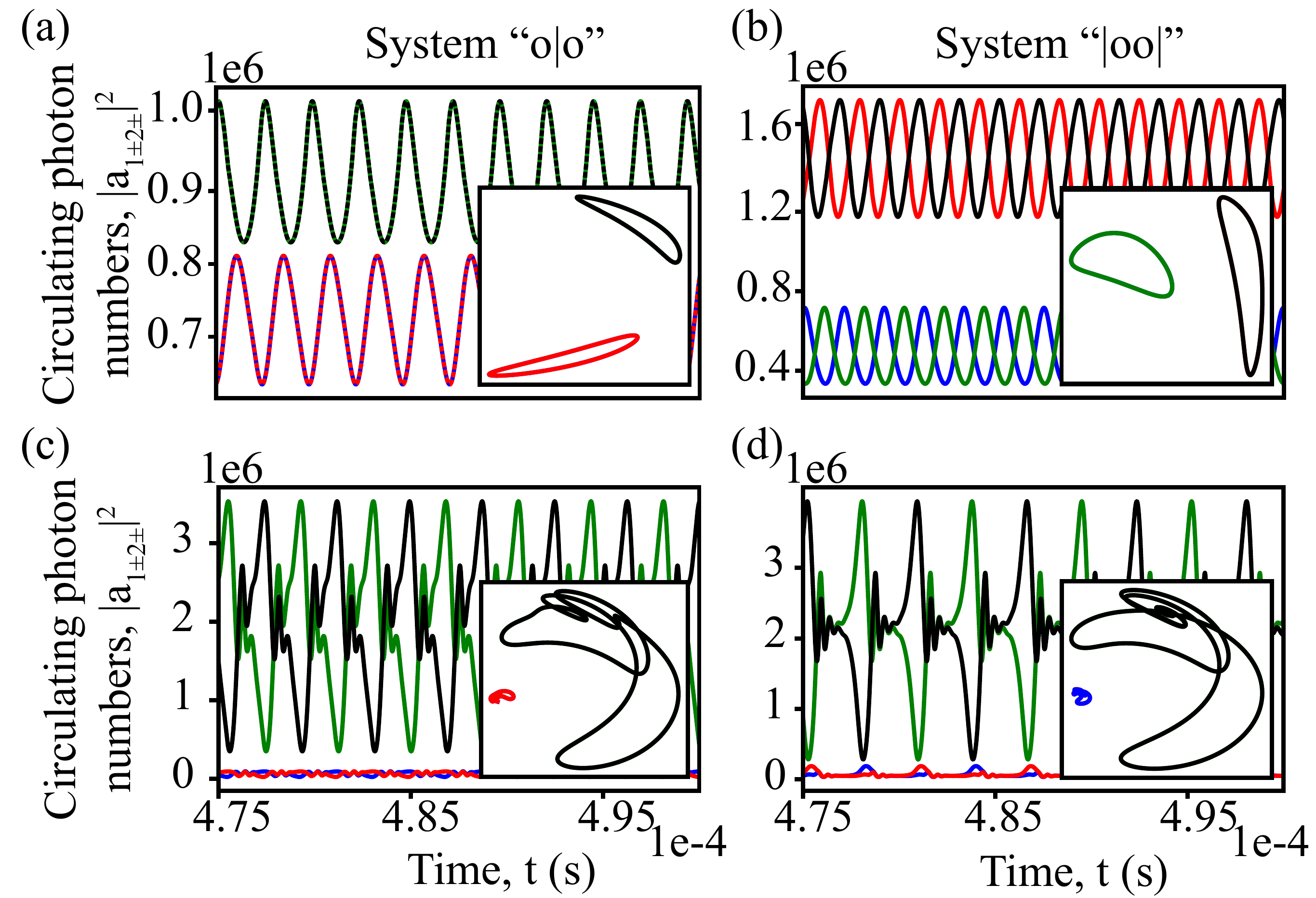}
\caption {\textit{Types of switching in the system.} Evolutions of field intensities over time in system ``o$\arrowvert$o" (a and c) and system ``$\arrowvert$oo$\arrowvert$" (b and d). In sinusoidal oscillatory regions where oscillations of any two pairs of fields overlap in system ``o$\arrowvert$o", the phases of the overlapping fields are same, as in (a) where $\Delta = -3.2 \kappa_l$ and $|S_{in}|^2 = 5.35 \times 10^{12}$. (b) Perfect sinusoidal switching of the cross fields ($|a_{1+}|^2$ with $|a_{2-}|^2$ and $|a_{1-}|^2$ with $|a_{2+}|^2$) in system ``$\arrowvert$oo$\arrowvert$" for $\Delta = -3.1 \kappa_l$ and $|S_{in}|^2 = 5.95 \times 10^{12}$. (c) and (d) show switching of fields within each resonator ($|a_{1+}|^2$ with $|a_{1-}|^2$ and $|a_{2+}|^2$ with $|a_{2-}|^2$) in system ``o$\arrowvert$o" and system ``$\arrowvert$oo$\arrowvert$" respectively. For (c), $\Delta = -6.66 \kappa_l$ and $|S_{in}|^2 = 1.29 \times 10^{13}$ and for (d), $\Delta = -7.17 \kappa_l$ and $|S_{in}|^2 = 1.31 \times 10^{13}$.}
\label{fig6}
\end{figure}

\section{Oscillations}
In Fig~\ref{fig6} we display different types of the self-switching oscillations~\cite{PhysRevLett.126.043901, PhysRevLett.128.053901} observed in the two systems. In system ``o$\arrowvert$o", sinusoidal field behavior is always accompanied by in phase oscillation of the pairing component as shown in Fig.~\ref{fig6}a. However, Fig.~\ref{fig6}b shows the self-switching oscillations between fields of two different resonators with mutually perpendicular polarizations in system ``$\arrowvert$oo$\arrowvert$". This switching is observed for $\Delta = -3.1 \kappa_l$, $|S_{in}|^2 = 5.95 \times 10^{12}$ and can be seen in a wide range of parameter values around this point. In Fig.~\ref{fig6}c and d, switching between the fields of different polarizations within the same resonators are plotted for system ``o$\arrowvert$o" and ``$\arrowvert$oo$\arrowvert$" respectively. The insets in (a)-(d) cases show perfect overlapping of the switching fields in phase space implying some global symmetry has been restored. One interesting phenomenon observed in the figures in the lower panels of Fig.~\ref{fig6} is that although switching of the different polarization components within both the resonators are observed in both systems, fields in one resonator get highly enhanced and in the other greatly suppressed.

In summary, we developed a theoretical framework to analyse the SSB of light in coupled twin resonators also known as photonic dimers. In the ``o$\arrowvert$o" photonic dimer system, two different kinds of 2-staged SSB have been observed, the symmetry breaking between the resonators and, for higher input intensities, the symmetry breaking between the cross pairs (one polarization of one resonator pairing with the orthogonal polarization of the other resonator). On the other hand, in the coupled photonic dimer, one extra type of 2-D symmetry breaking has been observed, which breaks the symmetry between the field polarizations. Full asymmetry between circulating photon numbers is accessible in both systems for relatively higher values of detuning. We found distinct regions of oscillations present in both the systems, each of which contains oscillations of fields with different orders of magnitude. The most interesting oscillations present in the systems were chaos and multiple variations of perfect periodic switching. In the geometrically uncoupled photonic dimer, we observe perfect periodic switching between the fields in the same resonators. In the ``$\arrowvert$oo$\arrowvert$" photonic dimer however we observed sinusoidal switching between the fields with same polarizations. Future works will address the effect of the loss terms and the inter-resonator coupling parameter on the stationary response of the system. This work will find applications in designing efficient Kerr-effect based polarization controllers, all optical computing, and designing compact optical isolators for quantum computers. This model further has the possibility of observing symmetry broken vector solitons with four different levels ~\cite{xu2021spontaneous}, which would be useful for generating four distinct frequency combs and would be very much useful in telecommunications and especially in space technologies due to compactness. 

 \section*{Acknowledgments}
LH acknowledges funding provided by the CNQO group within the Department of Physics at the University of Strathclyde, and the ``Saltire Emerging Researcher" scheme through the Scottish University's Physics Alliance (SUPA) and provided by the Scottish Government and Scottish Funding Council. This work was further supported by the European Union’s H2020 ERC Starting Grants 756966, the Marie Curie Innovative Training Network “Microcombs” 812818 and the Max Planck Society.

A.G. and L.H. contributed equally to this work.


\bibliography{Refs}

\onecolumngrid

\appendix
\section{Steady state responses}
\label{A}
The Kerr effect in each of the resonators can be described by the Hamiltonian, $\hat{H}_{j}^{Kerr} = -\left(\hbar/2\right)U\hat{a}_j^{\dagger}\hat{a}_j^{\dagger}\hat{a}_j\hat{a}_j$ where $\hbar$ is the Planck constant, $\hat{a}_j^{\dagger}(\hat{a}_j)$ is the creation (annihilation) operator in the $j^{th}$ resonator ($j \in \{1,2\}$), such that, $[\hat{a}_j, \hat{a}_k^{\dagger}] = \delta_{jk}$ and $[\hat{a}_j, \hat{a}_k] = 0$. The term $U = \frac{\hbar \omega_0^2 c n_2}{n^2 V_{\text{eff}}}$~\cite{PhysRevA.97.013843, Tikan2021, PhysRevB.85.033303} is the Kerr coefficient, where $\omega_0$ is the excited pump frequency in the microresonator, $c$ is the speed of light, $n_2$ is the nonlinear refractive index, $n$ is the linear refractive index and $V_{\text{eff}}$ is the effective mode volume. If we consider right and left handed circularly polarized fields inside the resonators, with creation (annihilation) operators $\hat{a}_{i\pm}^{\dagger} (\hat{a}_{i\pm})$, the Hamiltonian can be written as~\cite{doi:10.1080/09500349114551711}, 

\begin{eqnarray}
\hat{H}_{j}^{Kerr} = -\frac{\hbar}{2}U\left((\hat{a}_{j+}^{\dagger})^2\hat{a}_{j+}^2 + (\hat{a}_{j-}^{\dagger})^2\hat{a}_{j-}^2 + 4d\hat{a}_{j+}^{\dagger}\hat{a}_{j+}\hat{a}_{j-}^{\dagger}\hat{a}_{j-}\right),
\label{kerrHamiltonian}
\end{eqnarray}
where $2d = 1+\left(\chi_{xxyy}/\chi_{xyxy}\right)$ and $\chi_{xxyy}$ and $\chi_{xyxy}$ are the nonlinear susceptibility tensor terms of the medium.
Therefore, one can write the self- and cross-phase modulation terms in the evolution equations of the operators as,
\begin{eqnarray}
\label{OperatorEquationsKerr}
\dot{\hat{a}}_{j\pm} = iU\left(\hat{a}_{j\pm}^{\dagger}\hat{a}_{j\pm} + 2d\hat{a}_{j\mp}^{\dagger}\hat{a}_{j\mp}\right)\hat{a}_{j\pm}.
\end{eqnarray}

Therefore the evolution equations of the four fields in the two resonators can be written as,
\begin{subequations}
\label{LangevinEquationsr}
\begin{eqnarray}
\dot{a}_{1+} &=& \left(i\Delta -\frac{\kappa}{2}\right) a_{1+} +\zeta a_{2+} + i U |a_{1+}|^2 a_{1+} + i 2U |a_{1-}|^2 a_{1+} + \sqrt{\kappa_e}S_{in},\\
\dot{a}_{1-} &=& \left(i\Delta -\frac{\kappa}{2}\right) a_{1-} +\zeta a_{2-} + i U |a_{1-}|^2 a_{1-} + i 2U |a_{1+}|^2 a_{1-} + \sqrt{\kappa_e}S_{in},\\
\dot{a}_{2+} &=& \left(i\Delta -\frac{\kappa}{2}\right) a_{2+} +\zeta a_{1+} + i U |a_{2+}|^2 a_{2+} + i 2U |a_{2-}|^2 a_{2+} + \sqrt{\kappa_e}S_{in},\\
\dot{a}_{2-} &=& \left(i\Delta -\frac{\kappa}{2}\right) a_{2-} +\zeta a_{1-} + i U |a_{2-}|^2 a_{2-} + i 2U |a_{2+}|^2 a_{2-} + \sqrt{\kappa_e}S_{in},
\end{eqnarray}
\end{subequations}
where $a_{1\pm, 2\pm}$ stands for the real-valued classical amplitudes of the optical modes. The term $\zeta$ depends upon the mechanism of coupling as mentioned in the main text. In both the cases we consider that fields in the two resonators with same polarization orientation are coupled and there is no cross coupling between fields with orthogonal polarization orientation in the two resonators. In system ``o$\arrowvert$o", where there is no geometric coupling between the two resonators, the fields within the two resonators are related through the input-output relations. The field in one resonator is coupled to the modes in the tapered fiber and those modes are coupled to the resonant mode in the other resonator. After a detailed calculation, it can be derived that, $\zeta = -\left(\kappa_e/2\right)$.
In system ``$\arrowvert$oo$\arrowvert$", where the fields within the two resonators geometrically overlap, the interaction between the optical modes are modelled by the interaction Hamiltonian, $\hat{H}^{int} = -\hbar J (\hat{a}_{1+}^{\dagger}\hat{a}_{2+} + \hat{a}_{2+}^{\dagger}\hat{a}_{1+} + \hat{a}_{1-}^{\dagger}\hat{a}_{2-} + \hat{a}_{2-}^{\dagger}\hat{a}_{1-})$. The term $J$ defines the coupling strength between the two resonators. This Hamiltonian leads to $\zeta = iJ$ in Eqs.~\eqref{LangevinEquationsr} for system ``$\arrowvert$oo$\arrowvert$".\\
\indent In steady state, $\dot{a}_{1+} = \dot{a}_{1-} = \dot{a}_{2+} = \dot{a}_{2-} = 0$, i.e.,
\begin{subequations}
\begin{eqnarray}
\label{LangevinEquationsSS}
\left(i\Delta -\frac{\kappa}{2}\right) a_{1+} +\zeta a_{2+} + i U |a_{1+}|^2 a_{1+} + i 2U |a_{1-}|^2 a_{1+} + \sqrt{\kappa_e}S_{in} &=& 0,\\
\left(i\Delta -\frac{\kappa}{2}\right) a_{1-} +\zeta a_{2-} + i U |a_{1-}|^2 a_{1-} + i 2U |a_{1+}|^2 a_{1-} + \sqrt{\kappa_e}S_{in} &=& 0,\\
 \left(i\Delta -\frac{\kappa}{2}\right) a_{2+} +\zeta a_{1+} + i U |a_{2+}|^2 a_{2+} + i 2U |a_{2-}|^2 a_{2+} + \sqrt{\kappa_e}S_{in} &=& 0,\\
\left(i\Delta -\frac{\kappa}{2}\right) a_{2-} +\zeta a_{1-} + i U |a_{2-}|^2 a_{2-} + i 2U |a_{2+}|^2 a_{2-} + \sqrt{\kappa_e}S_{in} &=& 0.
\end{eqnarray}
\end{subequations}
Solving this system of equations is quite difficult when all the fields are asymmetric. Therefore, to study multi-staged symmetry breakings, leading to full asymmetry in the system, we let Eqs.~\eqref{LangevinEquationsr} evolve for a long time for increasing input power and record the final states. However, it is possible to study fully symmetric solution and different 2-staged symmetry breaking conditions in the system analytically. To do this, we impose the corresponding conditions of forced symmetry among different pairs of fields in the equations.


\subsection{Fully symmetric solution}
Here we consider, $a_{1+} = a_{1-} = a_{2+} = a_{2-} = a$. Therefore, Eqs.~\eqref{LangevinEquationsr} takes the form,
\begin{eqnarray}
\label{HSS}
\dot{a} &=& \left(i\Delta -\frac{\kappa}{2}\right) a+\zeta a+ i U |a|^2 a+ i 2U |a|^2 a + \sqrt{\kappa_e}S_{in}.
\end{eqnarray}
The steady state in this case can be described as,
\begin{subequations}
\label{HSS_Poly}
\begin{align}
A^3(9U^2) &+ A^2(6\Delta U) + A \left(  \Delta^2 + \frac{9 \kappa_e^2}{4} \right) - \kappa_e|S_{in}|^2 = 0 \text{, for system ``o$\arrowvert$o"},\\
A^3(9U^2) &+ A^2 6U (\Delta + J) + A \left(  \Delta^2 + J^2 + 2\Delta J + \frac{\kappa^2}{4} \right) - \kappa_e|S_{in}|^2 = 0 \text{, for system ``$\arrowvert$oo$\arrowvert$"},
\end{align}
\end{subequations}
where $A = |a|^2$.


\subsection{Polarization Symmetry (PS)}
Here we consider, $a_{1+} = a_{1-} = b$ and $a_{2+} = a_{2-} = c$. Therefore, Eqs.~\eqref{LangevinEquationsr} takes the form,
\begin{subequations}
\label{RSB}
\begin{align}
\dot{b} &= \left\{i\left(\Delta + 3U|b|^2\right) -\frac{\kappa}{2}\right\} b+\zeta c + \sqrt{\kappa_e}S_{in},\\
\dot{c} &= \left\{i\left(\Delta + 3U|c|^2\right) -\frac{\kappa}{2}\right\} c+\zeta b + \sqrt{\kappa_e}S_{in}.
\end{align}
\end{subequations}
The steady state in this case can be described as,
\begin{subequations}
\label{RSB_Poly}
\begin{align}
B^3(9U^2) &+ B^2(6\Delta U) + B \left(  \Delta^2 + \frac{\kappa_e^2}{4} \right) - \left( \Delta^2 C + 9 U^2 C^3 + 6\Delta U C^2 + \frac{\kappa_e^2}{4}C \right) = 0 \text{, for system ``o$\arrowvert$o"},\\
\nonumber
B^3(9U^2) &+ B^2 6U (\Delta - J) + B \left(  \Delta^2 + J^2 - 2\Delta J + \frac{\kappa^2}{4} \right)\\
&- \left( \Delta^2 C + 9U^2C^3 + J^2C + 6\Delta U C^2 - 2 \Delta J C - 6UJC^2 + \frac{\kappa^2}{4}C \right) = 0 \text{, for system ``$\arrowvert$oo$\arrowvert$"},
\end{align}
\end{subequations}
where $B = |b|^2$ and $C = |c|^2$.


\subsection{Resonator symmetry (RS)}
Here we consider, $a_{1+} = a_{2+} = d$ and $a_{1-} = a_{2-} = e$. Therefore, Eqs.~\eqref{LangevinEquationsr} takes the form,
\begin{subequations}
\label{RS}
\begin{eqnarray}
\dot{d} &=& \left\{i\left(\Delta + U|d|^2 + 2U|e|^2\right) -\frac{\kappa}{2}\right\} d+\zeta d + \sqrt{\kappa_e}S_{in},\\
\dot{e} &=& \left\{i\left(\Delta + U|e|^2 + 2U|d|^2\right) -\frac{\kappa}{2}\right\} e+\zeta e + \sqrt{\kappa_e}S_{in}.
\end{eqnarray}
\end{subequations}

The RS solution is only observed in case of system ``$\arrowvert$oo$\arrowvert$". In that case, the steady state can be described as,
\begin{align}
\label{RS_Poly}
\nonumber
D^3(U^2) + D^2(2\Delta U+2UJ) &+ D \left(  \Delta^2 + \frac{\kappa^2}{4} + J^2 + 2\Delta J \right)\\
 &- \left\{ E^3 U^2 + E^2 \left( 2\Delta U + 2 U J \right) + E\left( \Delta^2 + J^2 + 2\Delta J + \frac{\kappa^2}{4} \right) \right\} = 0,
\end{align}
where $D = |d|^2$ and $E = |e|^2$.


\subsection{Cross symmetry (CS)}
Here we consider, $a_{1+} = a_{2-} = f$ and $a_{1-} = a_{2+} = g$. Therefore, Eqs.~\eqref{LangevinEquationsr} takes the form,
\begin{subequations}
\label{XSB}
\begin{eqnarray}
\dot{f} &=& \left\{i\left(\Delta + U|f|^2 + 2U|g|^2\right) -\frac{\kappa}{2}\right\} f+\zeta g + \sqrt{\kappa_e}S_{in},\\
\dot{g} &=& \left\{i\left(\Delta + U|g|^2 + 2U|f|^2\right) -\frac{\kappa}{2}\right\} g+\zeta f + \sqrt{\kappa_e}S_{in}.
\end{eqnarray}
\end{subequations}
The steady state in this case can be described as,
\begin{subequations}
\label{XSB_Poly}
\begin{align}
F^3(U^2) + 2\Delta U F^2 &+ F \left(  \Delta^2 + \frac{\kappa_e^2}{4} \right) - \left( \Delta^2 G + U^2 G^3 + 2\Delta U G^2 + \frac{\kappa_e^2}{4}G \right) = 0 \text{, for system ``o$\arrowvert$o"},\\
\nonumber
F^3(U^2) + F^2(2\Delta U&-2UJ) + F \left(  \Delta^2 + \frac{\kappa^2}{4} + J^2 - 2\Delta J \right)\\
 &- \left\{ G^3 U^2 + G^2 \left( 2\Delta U - 2 U J \right) + G\left( \Delta^2 + J^2 - 2\Delta J + \frac{\kappa^2}{4} \right) \right\} = 0  \text{, for system ``$\arrowvert$oo$\arrowvert$"},
\end{align}
\end{subequations}
where $F = |f|^2$ and $G = |g|^2$.
\end{document}